\documentclass{PoS}

\usepackage{txfonts}
\usepackage{booktabs}
\usepackage{graphicx}
\usepackage{threeparttable}

\title{A low-redshift low luminosity QSO sample: Comparison with NUGA galaxies and PG QSOs and first interferometric images of three sample members}

\ShortTitle{A LLQSO sample: Comparison with NUGA galaxies and PG QSOs}

\author{\speaker{Lydia Moser} $^a$, Jens Zuther $^a$, Sebastian Fischer $^a$, Gerold Busch $^a$, Monica Valencia-S.$^a$, 
Andreas Eckart$^{a,b}$, Melanie Krips $^c$, Sabine K\"onig $^c$, Julia Scharw\"achter $^d$ 
\\
        \llap{$^a$} 1. Physikalisches Institut, University of Cologne, Germany\\
        \llap{$^b$} Max-Planck-Institut f\"ur Radioastronomie, Germany\\
        \llap{$^c$} Institut de Radioastronomie Millim\'etrique (IRAM), Grenoble, France\\
        \llap{$^d$} Observatoire de Paris, Paris, France\\
        E-mail: \email{moser@ph1.uni-koeln.de}, \email{zuther@ph1.uni-koeln.de}, \email{fischer@ph1.uni-koeln.de}, \email{busch@ph1.uni-koeln.de}, \email{mvalencias@ph1.uni-koeln.de}, \email{eckart@ph1.uni-koeln.de}, \email{krips@iram.fr}, \email{koenig@iram.fr}, \email{julia.scharwaechter@obspm.fr} 
        }

\abstract{
The low luminosity QSO (LLQSO) sample consists of type 1 active galactic nuclei (AGN) up to a redshift of $z=0.06$ in the Hamburg/ESO QSO survey. Its purpose is to study how the brightest AGN in the nearby universe evolve with respect to AGN activity and host properties as a function of redshift. We show that our sample lies well between the NUclei of GAlaxies (NUGA) sample and the Palomar Green (PG) QSO sample in terms of redshift, gas masses and luminosities and seems to connect them. The continuous growth in mass, luminosity and, linked to this, the AGN activity over the samples has either a statistical reason or is indicative of an evolutionary link between the different populations and might be related to cosmic downsizing. 
In addition, we present first results of our observations of three galaxies from our sample with the Submillimeter Array (SMA).  
}

\FullConference{Nuclei of Seyfert galaxies and QSOs - Central engine \& conditions of star formation\\
		 November 6-8, 2012\\
		 Max-Planck-Institut f\"ur Radioastronomie (MPIfR), Bonn, Germany}

\begin{document}

\section{Introduction}

The strong correlation between the black hole mass and the bulge velocity dispersion of galaxies suggests a coeval growth of supermassive black holes (SMBH) and the surrounding stellar bulges. The growth is believed to be regulated by an interplay of nuclear fueling, i.e. inflow of gas which is then consumed by star formation and accretion onto the SMBH, and feedback from these regions by winds, outflows and radiation. The diverse mechanisms that might be involved in these processes (e.g. galactic interactions or secular processes) are topics of current research. While for high-luminosity active galactic nuclei (AGN) the gas inflow is likely to be triggered by large-scale bars or major galaxy interactions (see e.g. [1], [2])
, secular evolution seems to be the dominating fueling mechanism for low-luminosity AGN (LLAGN, see e.g. [3], [1], [4]).

The goal of our low luminosity QSO (LLQSO) sample is to investigate the nuclear fueling in AGN more powerful than the local LLAGN in order to study their relation to the local population and the more active galaxies at higher redshifts.

The sample 
([1]) has been selected from the Hamburg/ESO QSO survey (flux limit of $B_J\leq 17.3$, 
[6]) with the selection criterion of $z\leq 0.06$ which yielded 99 sources. Up to this redshift limit the CO(2--0) band head from stellar atmospheres in the near--infrared (NIR) $K$--band is observable so that stellar and AGN properties can be analyzed.
Furthermore, the sub-kpc scale can still be sufficiently resolved by the current telescopes and arrays to disentangle the starburst from the AGN component and to investigate the interstellar medium (ISM) in the circumnuclear environment.

In fact, the Atacama Large Millimeter Array (ALMA) enables us to observe the LLQSOs with spatial scales (few 10 pc) comparable to those achieved for the NUGA sources (NUclei of GAlaxies, see e.g. [7]) 
with the Plateau de Bure Interferometer (PdBI) and with a signal-to-noise ratio (SNR) for the high density tracers of the order of the SNR obtained for NUGA submillimeter observations of CO. 
In terms of redshift and activity range our sample adds to the NUGA survey and enlarges the volume from the local to the nearby universe as well as the part of the AGN activity sequence that can be studied in detail to higher activity levels, that are typical for e.g. the Palomar Green (PG) QSO sample ([8], [9], [10]).
The PG QSOs are already at a redshift at which the ALMA could only resolve them on few 100 pc scales.

40\% of the LLQSO sample has already been observed in the CO(1--0) and (2--1) transition with the IRAM 30~m telescope and with a detection rate of 70\% ([5]).
This paper compares some redshift dependent global properties of our CO-detected HE sources with the NUGA and the PG QSO sample. In addition, we present first results from an interferometric observation of CO line emission in three galaxies from our sample with the Submillimeter Array (SMA).

\begin{figure}[hptb]                                                     
\centering $                                                            
\begin{array}{cc}             
\includegraphics[trim = 0mm 20mm 20mm 30mm, clip, width=0.46\textwidth]{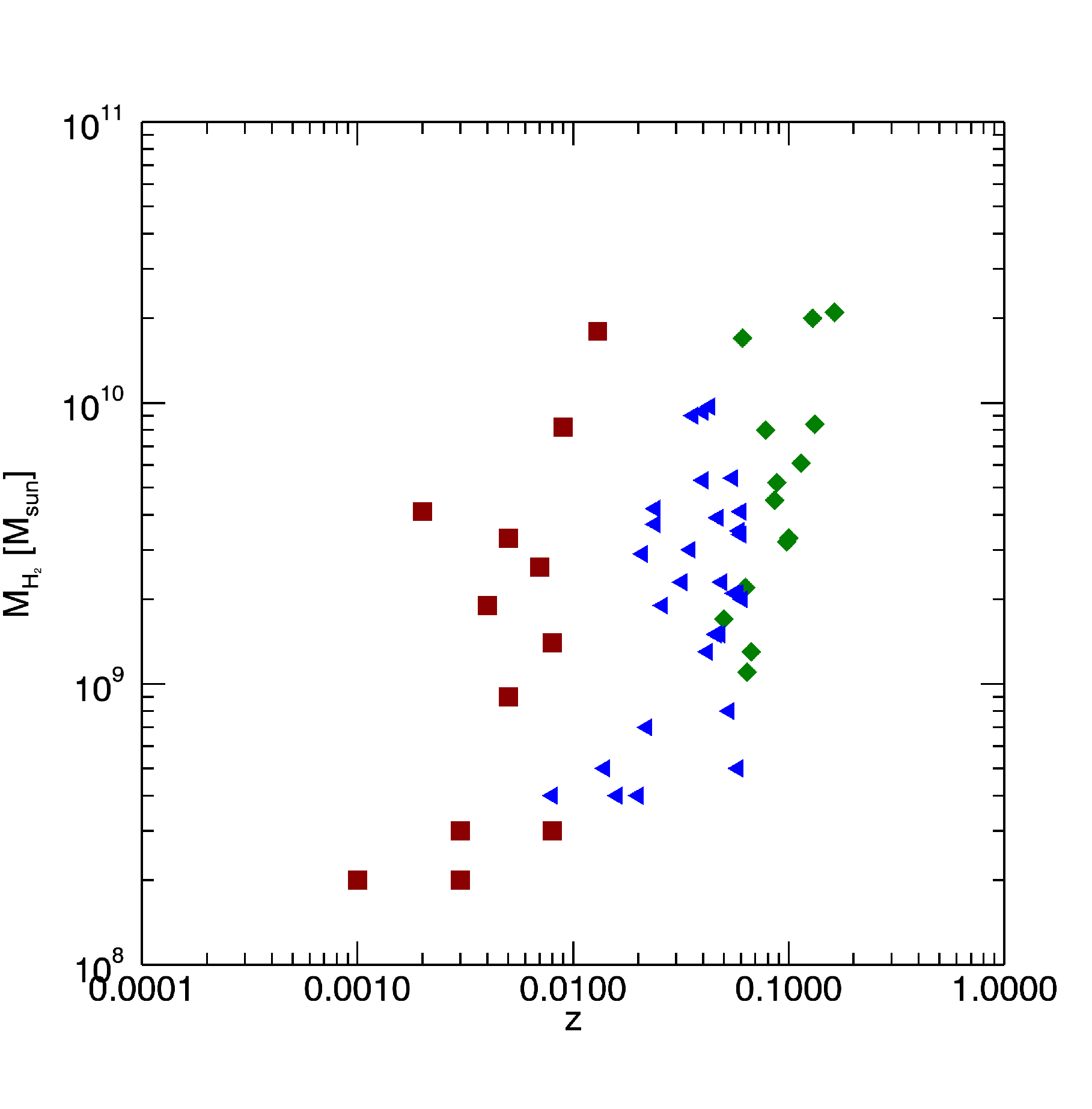}&
\includegraphics[trim = 0mm 20mm 20mm 30mm, clip, width=0.46\textwidth]{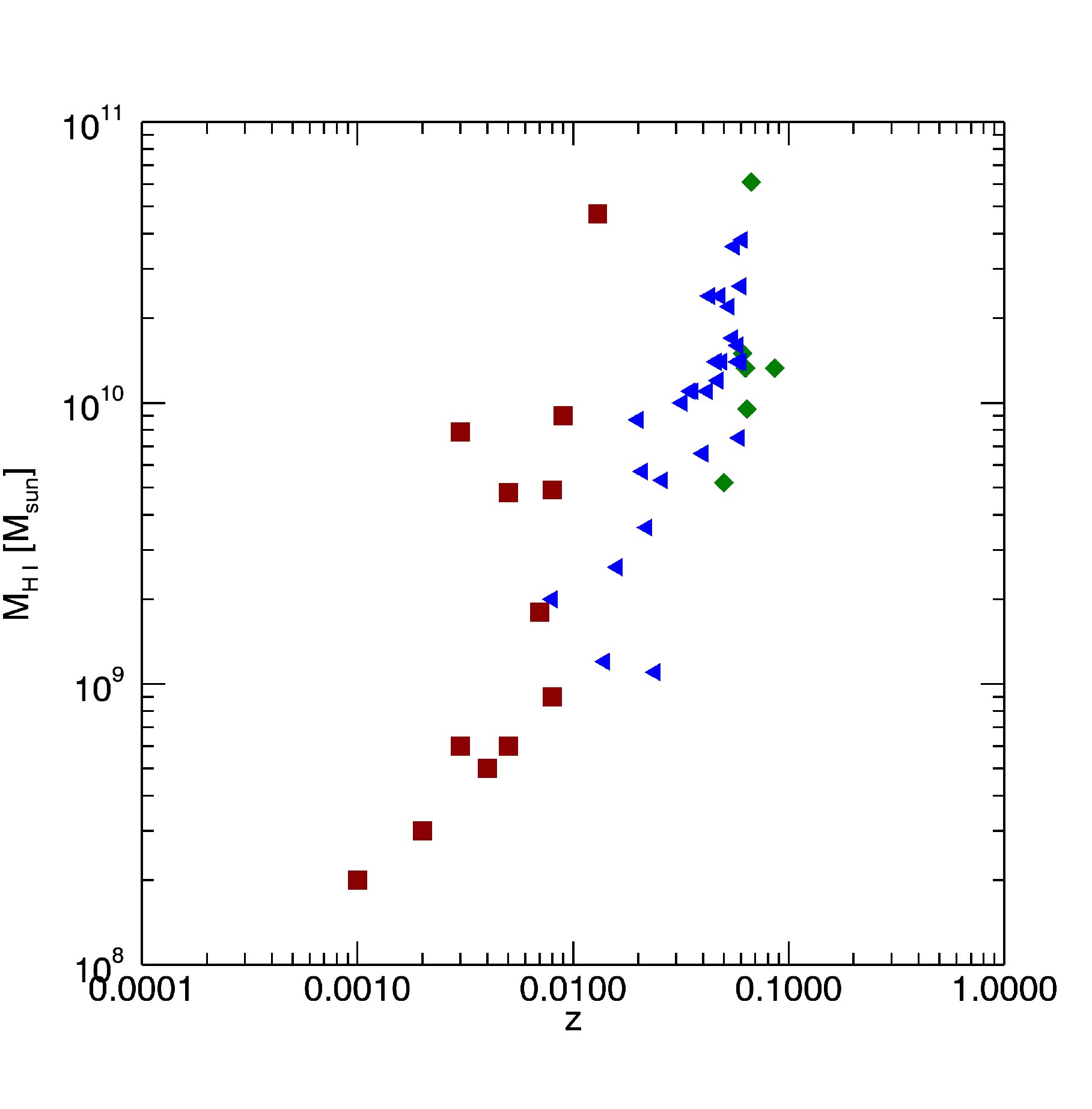}\\
\includegraphics[trim = 0mm 20mm 20mm 30mm, clip, width=0.46\textwidth]{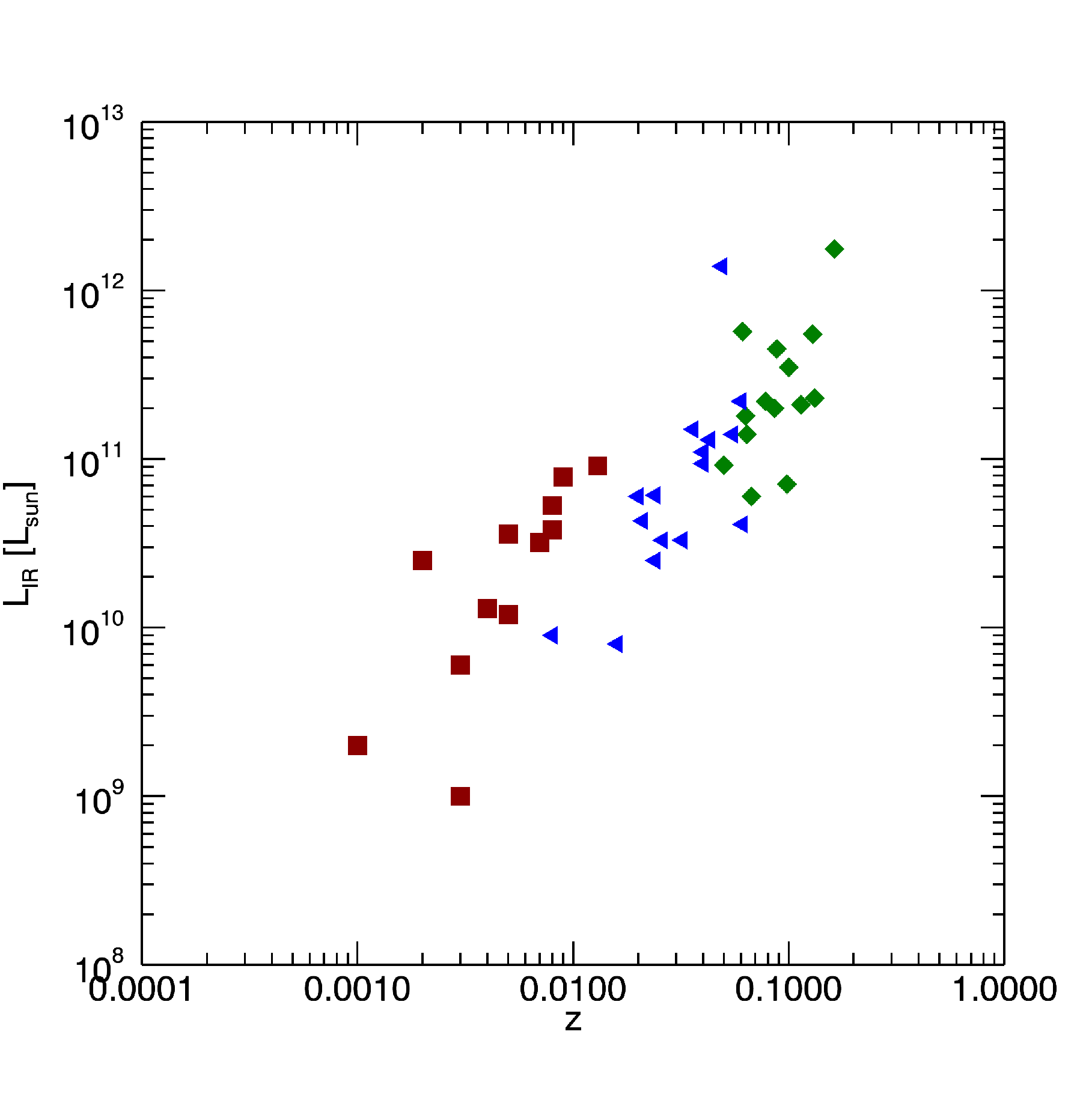}&
\includegraphics[trim = 0mm 20mm 20mm 30mm, clip, width=0.46\textwidth]{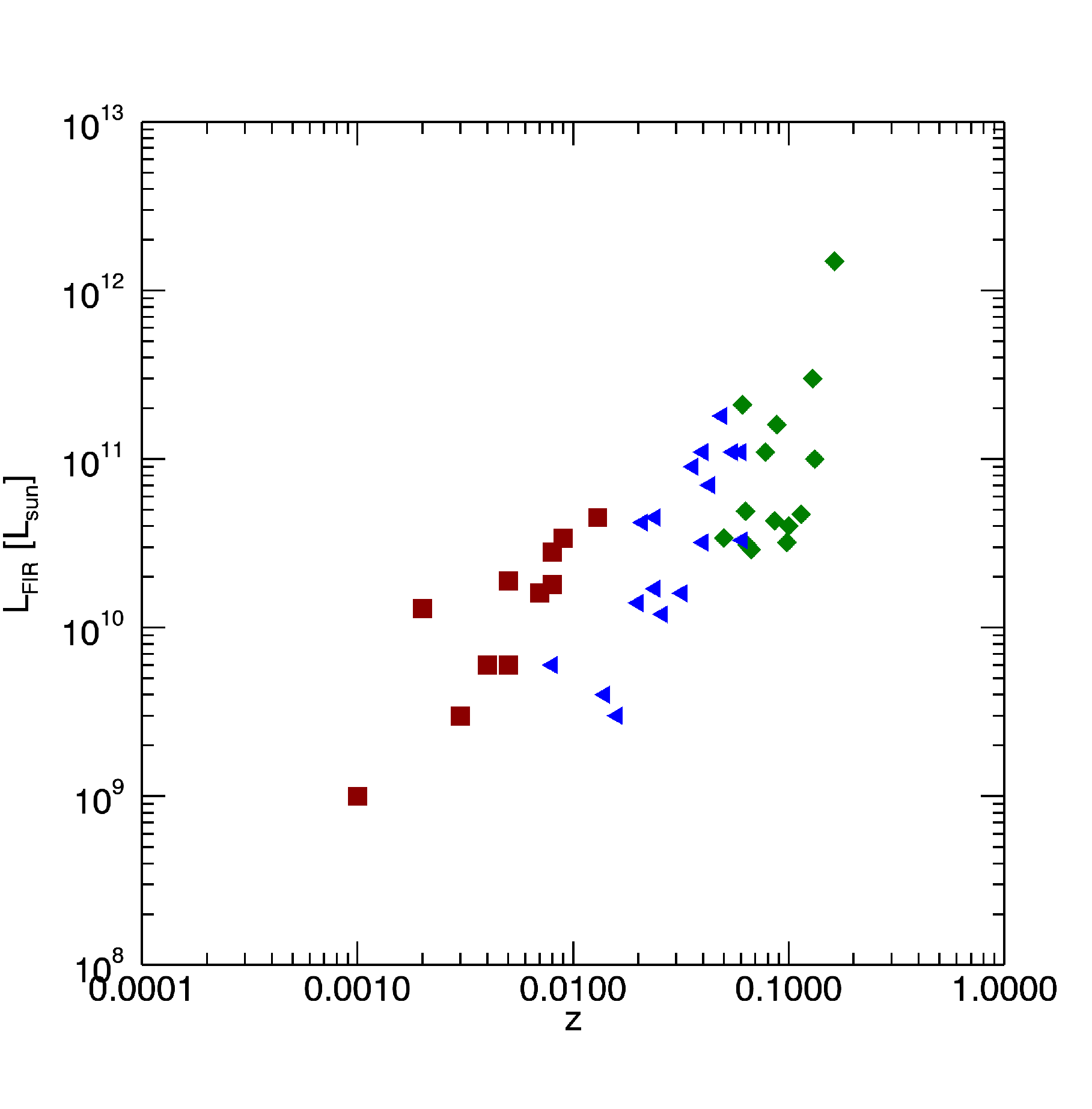}\\
\includegraphics[trim = 0mm 20mm 20mm 30mm, clip, width=0.46\textwidth]{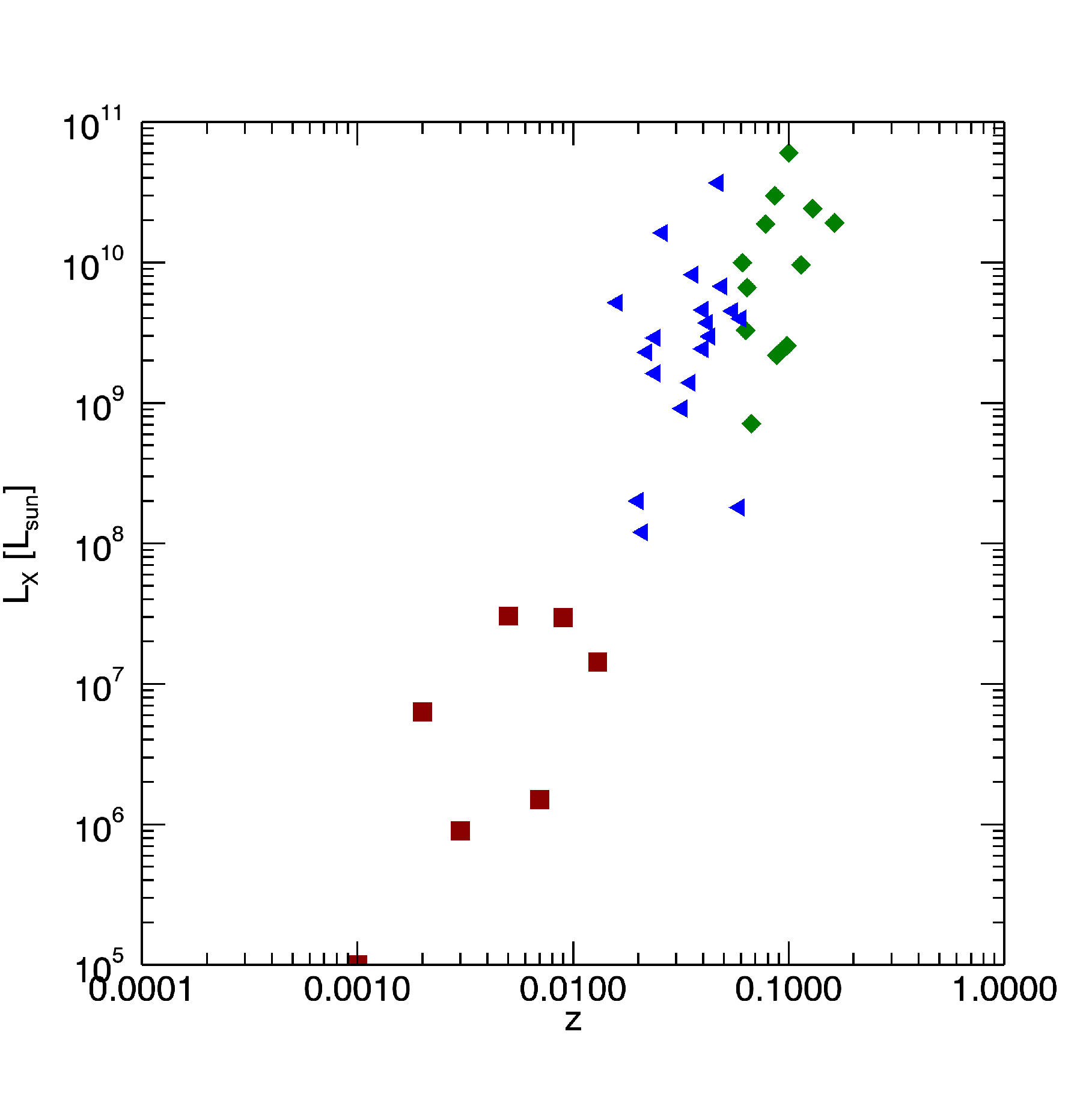}&
\includegraphics[trim = 0mm 20mm 20mm 30mm, clip, width=0.46\textwidth]{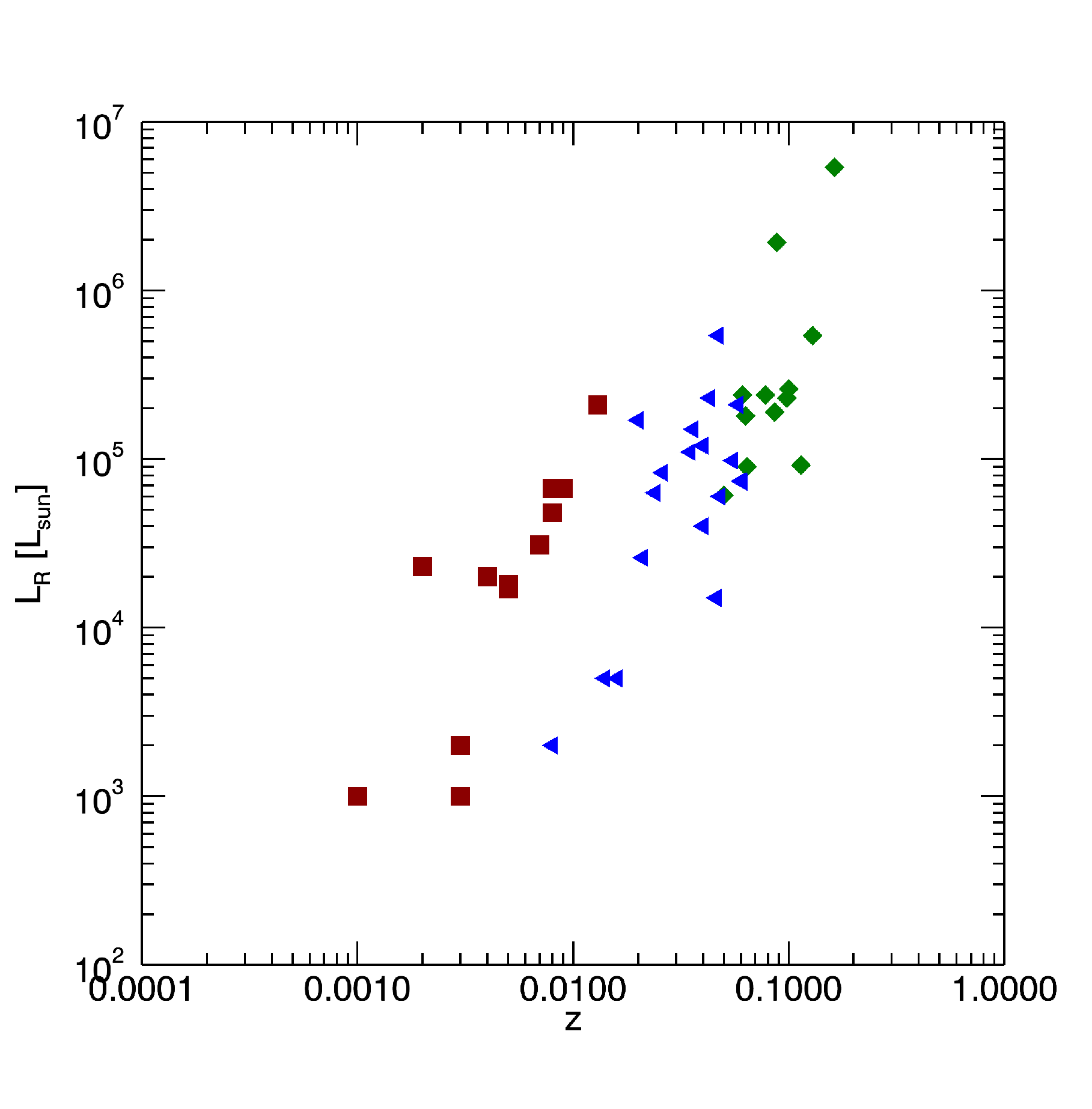}\\
\end{array} $
\caption{Different properties of the NUGA sources (red squares), CO-detected LLQSOs (HE sources, blue triangles) and PG QSOs (green diamonds) plotted against the redshift. From left to right and top to bottom: \emph{molecular gas mass} (NUGA: [7], [11], [2], [13], [14], [15], [16], [17], [18], [19], [20], [21];
 HE: [5], [22];
 PG QSOs: [8], [9], [10];
 common CO-conversion-factor $\alpha$ $\sim$ 4 M$_\odot$ [K km s$^{-1}$ pc$^2$]$^{-1}$), \emph{atomic gas mass} (NUGA: [23];
 HE: [24];
 PG QSOs: derived from H I fluxes (NED)), \emph{infrared} (IRAS-NED and formalism of [25]
for the 8-1000 micron range), \emph{far-infrared} (IRAS-NED and formalism of [26]
for the 40-120 micron range), \emph{X-ray} (0.5-2.0 keV flux (XMM, NED) or 0.1-2.4 keV flux (ROSAT, NED)) and \emph{radio} (1.4 GHz flux (NED)) \emph{luminosity} vs. redshift}
\label{vs-z}
\end{figure}

\section{Comparison to the NUGA and PG QSO sample}

Figure \ref{vs-z} visualizes that our CO-detected LLQSOs cover the apparent gap between the distributions of NUGA sources and PG QSOs not only in terms of redshift but also in terms of gas masses and luminosities.
The LLQSOs clearly follow the trend of increasing AGN and star forming activity with increasing redshift. 
There are two possible explanations for this behavior. First, the probability to detect a more powerful AGN and with this most likely a more massive supermassive black hole and central spheroid (Schechter-function) increases with the volume size. The samples cover volumes of different size, NUGA the smallest, PG QSOs the largest. Therefore, this trend can be explained as due to a volume/luminosity effect. 
Second, the trend might demonstrate the "cosmic downsizing" in the luminosity and mass distribution ([27], [28], [29], [30], [31], [32]).
In this scenario massive black holes and spheroids grow first, i.e. at high redshift, and low mass systems later, see e.g. local/nearby universe.  
Consequently, our sample would not only contribute to the study of the evolutionary connection between local and more distant LLAGN, but also to the test of the anti-hierarchical black hole growth hypothesis for the population of LLAGN in the nearby universe.
In order to shed light on these questions a statistically relevant sample of the order of 100 objects is essential.

\section{Interferometric data and ongoing work}

\begin{figure}[hptb]                                                     
\centering $                                                            
\begin{array}{ccc}             
\includegraphics[
height=0.23\textheight]{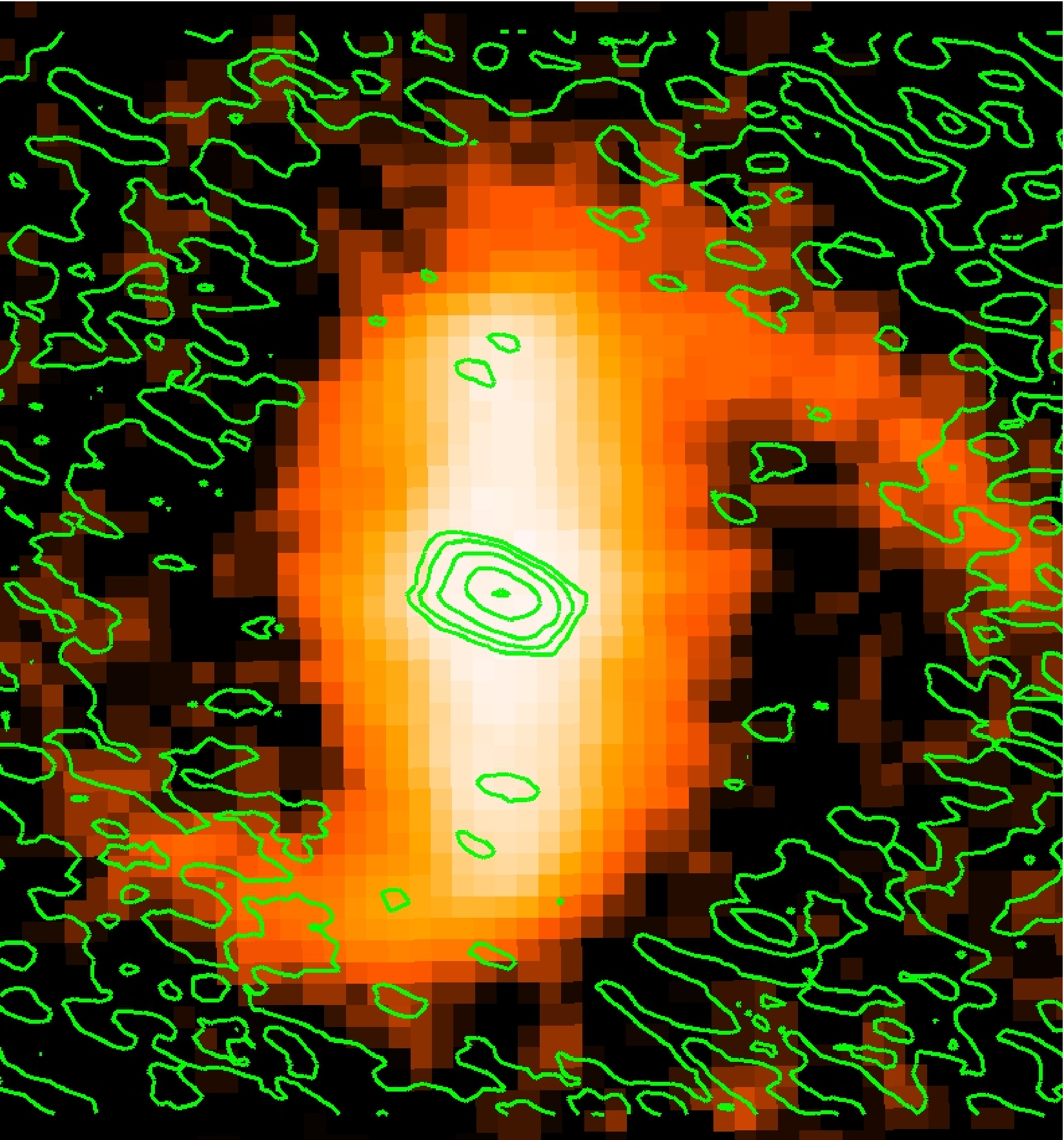}&
\includegraphics[
height=0.23\textheight]{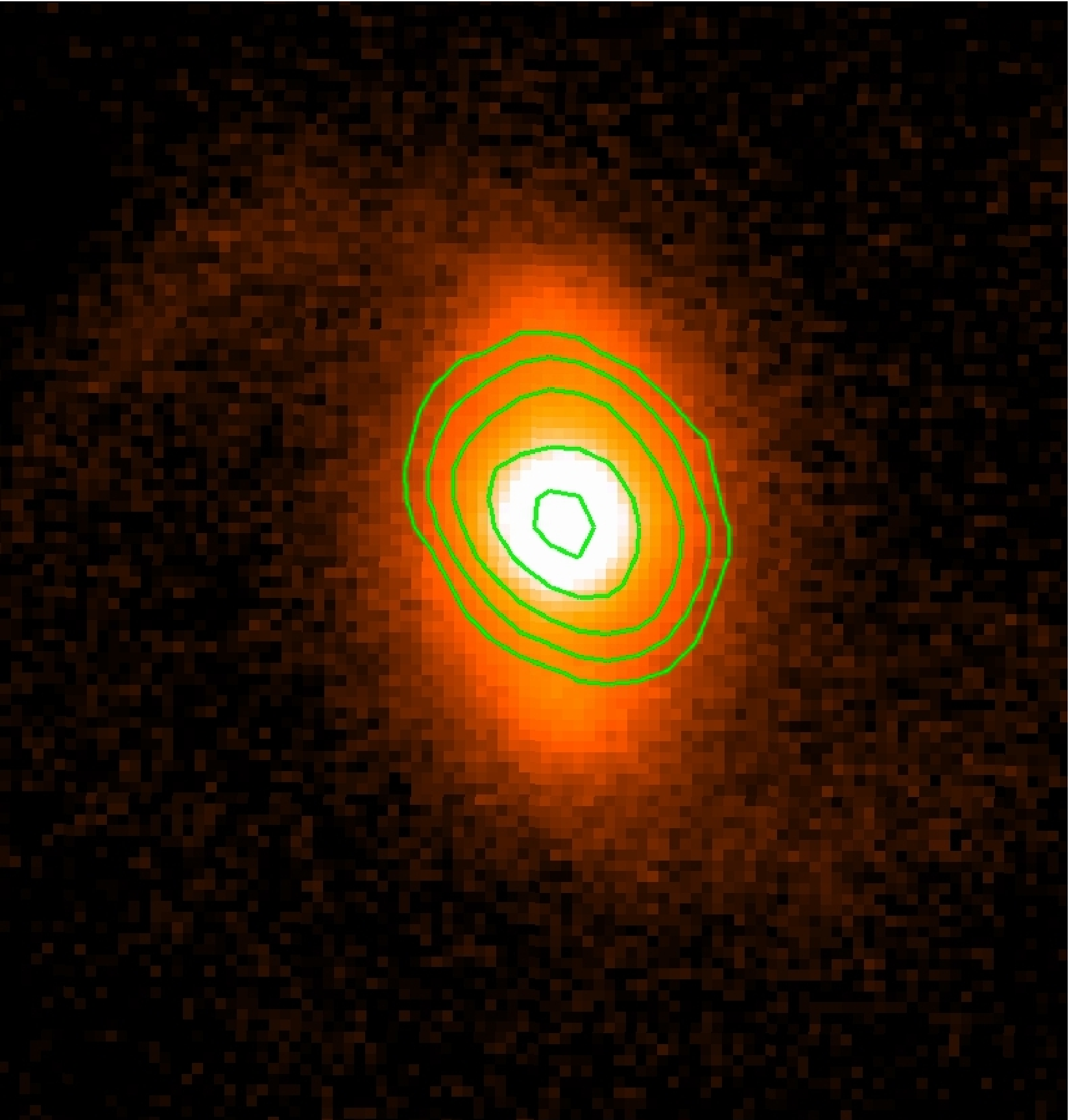}&
\includegraphics[trim = 0mm 15mm 0mm 20mm, clip, height=0.23\textheight]{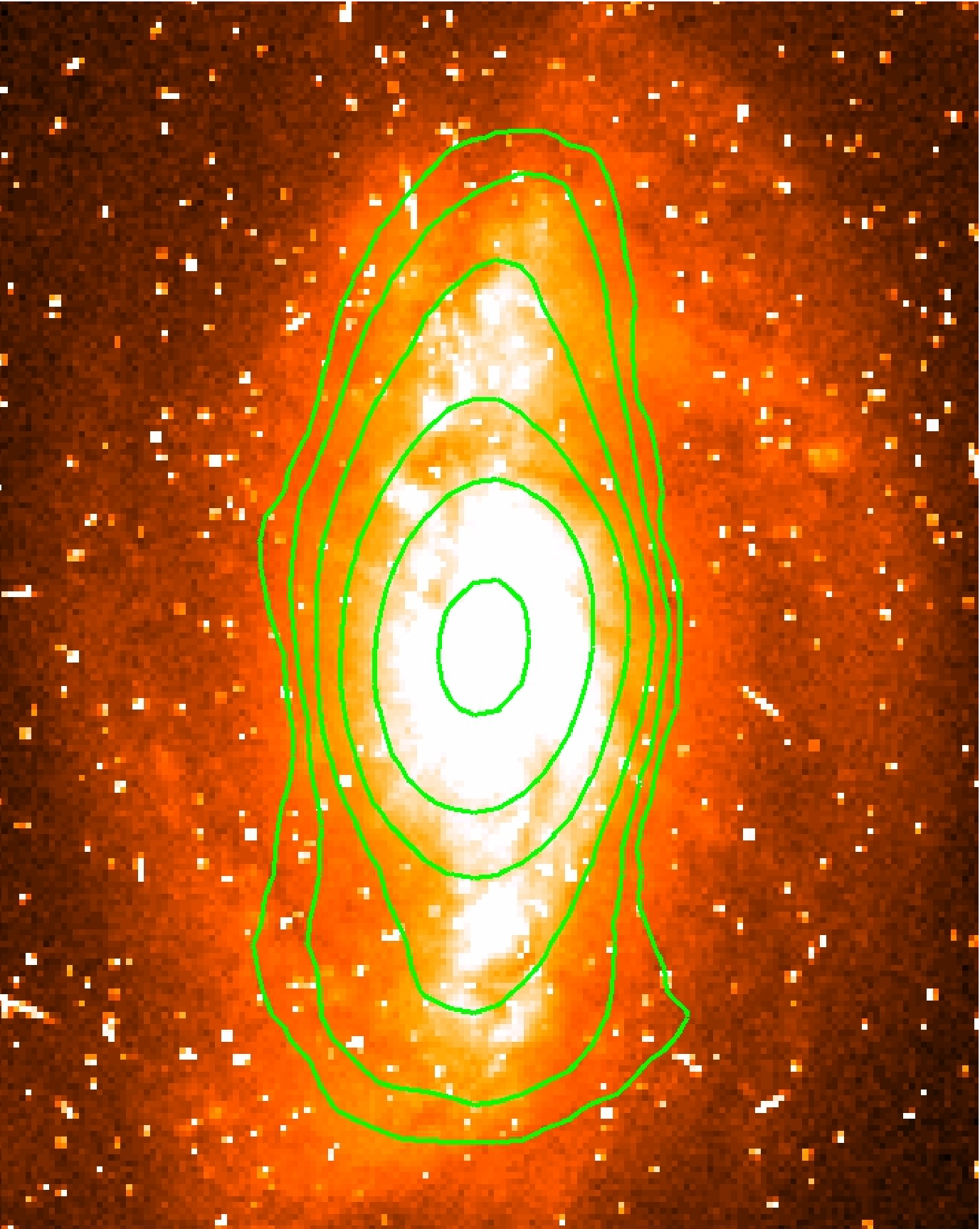}\\
\end{array} $
\caption{From left to right (based on naturally weighted visibilities): \emph{HE0433--1028}: the integrated CO(2--1) intensity map with contours in steps of (1, 2, 4, 8, 12) $\times$ 3$\sigma$  (= 2.7 JJy beam$^{-1}$ km s$^{-1}$) and a beam size of $3.9'' \times 1.6''$, overlayed onto an UK Schmidt image (red, DSS, noise at edges elevated due to primary beam correction); \emph{HE1029--1831}: the integrated CO(3--2) intensity map with contours in steps of (1, 2, 4, 8, 12) $\times$ 3$\sigma$  (= 6 Jy beam$^{-1}$ km s$^{-1}$) and a beam size of $2.5'' \times 1.5''$, overlayed onto an ISAAC image (H-band, Fischer, private communication); \emph{HE1108--2813}: the integrated CO(2--1) intensity map with contours in steps of (1, 2, 4, 8, 16) $\times$ 3$\sigma$  (= 2.1 Jy beam$^{-1}$ km s$^{-1}$) and a beam size of $4.5'' \times 3.0''$, overlayed onto an HST image (F606, red, HST archive).}
\label{HE-images}
\end{figure}

Three galaxies, belonging to the brightest detections in CO emission, have been observed with the SMA as a follow-up to our IRAM 30m single dish study. Currently, the CO data of these three galaxies is analyzed, soon to be published in Moser et al. 2013 (in prep., see \ref{HE-images}). 
We find the molecular gas with $M_{\mathrm{H}_2+\mathrm{He}} \sim 10^8 - 10^9 M_\odot$ to be concentrated within regions $\leq$ 1.6 kpc.
The emission region size as well as the consistency between the gas and dust mass when using a gas mass conversion factor and gas to dust ratio typical for ULIRGs indicate that these LLQSOs could be downsized versions of ULIRGs or high luminosity QSO. In terms of morphology the comparison between the SMA data and public optical/NIR images imply that HE 0433--1028 and HE 1108--2813 might contain a nested bar whereas HE1029--1831 seems to display a lopsidedness in the CO emission (for details, see Moser et al. 2013 (in prep.)).

Apart from this submillimeter work, the sample has been investigated with regard to many other aspects: First results of a near-infrared study of a sub-sample of our LLQSOs are presented in Busch et al. 2013 (these proceeding), Zuther et al. 2013 (these proceeding) investigated the NLS1 galaxies in our LLQSO sample and Tremou et al. 2013 (in prep.) discuss the optical spectra of the LLQSOs.

\end{document}